\documentclass{IEEEtran}
\usepackage{subfigure}
\usepackage{amsmath,epsfig,latexsym}
\usepackage{multicol}
\usepackage{bm}
\usepackage{subfigure}
\usepackage{amssymb}
\usepackage{amsfonts}
\usepackage{graphicx}
\usepackage{url}
\usepackage{ccaption}
\usepackage{booktabs}
\usepackage{color}
\usepackage{multirow}
\usepackage{amsthm}

\newtheorem{proposition}{Proposition}
\usepackage{algorithm}
\usepackage{algorithmic}

\newcommand{\ist}{\hspace*{.3mm}}

\ifCLASSINFOpdf
\else
\fi

\begin{document}
%

\title{Multi-sensor Suboptimal Fusion Student's $t$ Filter}

%

\author{Tiancheng~Li, Zheng Hu, Zhunga Liu and Xiaoxu Wang

\thanks{This work was supported by the National Natural Science Foundation of China under Grant 62071389.}
\thanks{
The authors are with the Key Laboratory of Information Fusion Technology (Ministry of Education), School of Automation, Northwestern Polytechnical University, Xi'an 710129, China, e-mail: t.c.li@nwpu.edu.cn (T.Li)
}}

\maketitle
\vspace{-1cm}
\begin{abstract}
A multi-sensor fusion Student's $t$ filter is proposed for time-series recursive estimation in the presence of heavy-tailed process and measurement noises. Driven from an information-theoretic optimization, the approach extends the single sensor Student's $t$ Kalman filter based on the suboptimal arithmetic average (AA) fusion approach. To ensure computationally efficient, closed-form $t$ density recursion, reasonable approximation has been used in both local-sensor filtering and inter-sensor fusion calculation. The overall framework accommodates any Gaussian-oriented fusion approach such as the covariance intersection (CI). Simulation demonstrates the effectiveness of the proposed multi-sensor AA fusion-based $t$ filter in dealing with outliers as compared with the classic Gaussian estimator, and the advantage of the AA fusion in comparison with the CI approach and the augmented measurement fusion.
\end{abstract}

\begin{IEEEkeywords}
Heavy-tailed noise, Student's $t$ filter, arithmetic average fusion, covariance intersection, multi-sensor fusion
\end{IEEEkeywords}

\IEEEpeerreviewmaketitle

\section{Introduction}
\IEEEPARstart{H}{eavy-tailed} noises are involved in many state estimation problems, e.g., tracking scenarios with agile/manoeuvering targets and outlier-corrupted measurements \cite{Roth17}. In these situations even when the state space model is linear, the performance of the predominant Kalman filter (KF) that models the noises as Gaussian often deteriorates. This gives rise to outlier-robust estimator based on heavy-tailed distribution such as the Student's $t$ for modelling the process and measurement noises \cite{Piche12,Zhu13VBStuT,Roth13,Huang16,Huang17}. The $t$ distribution can be viewed as a generalized Gaussian distribution that has an adjustable parameter referred to as the degree of freedom (dof) to uplift the tail of the distribution \cite{Nadarajah05}. As the key to drive the Bayes recursive Student's $t$ filter, the joint probability density function (PDF) of the state and process/measurement noises is assumed to be Student's $t$, and the prediction/posterior state PDF is then approximated as Student's $t$ \cite{Roth13,Roth17,Huang16} based on linear transformation of $t$ distribution. Different strategies have been further proposed for dealing with nonlinearity using such as the Monte Carlo method \cite{Genz98,Chai06tPF}, linearization \cite{Roth13}, and unscented transform \cite{Huang16,Tronarp16}. 

So far, the majority of existing Student's $t$ filters are implemented on the base of a single sensor except for few exceptions that seek optimal fusion of Student's $t$ in the sense of minimum variance (MV) estimation \cite{Chen18distributedStu,Yan21chapter}. First, these fusion approaches rely restrictively on accurate knowledge (or neglection) of the correlation between sensors (or to say, the common information among sensors \cite{Uhlmann96,Julier01,Bailey12}). Failure to properly account for the correction will lead to inconsistent estimation (e.g., underestimating the actual squared estimate errors) and even filter breakdown. 
Second, a key of the MV-oriented fusion is to get as accurate estimate as possible which prefers actually light-tailed density and is lack of congruity with the idea of the heavy-tailed estimator where the robustness-to-outliers stems from a heavy-tailed posterior. These observations motivate us to develop novel fusion approach that does not need to calculate the inter-sensor correlation or common information yet avoids inconsistent estimate and preserves the heavy tail of the $t$ estimator. Furthermore, when a large scale sensor network that is highly restrictive in both communication and computation is involved, computation efficiency and tolerance to node faults are also vital factors that need to be taken into account. %

In this paper, we propose a multi-sensor Student's $t$ filter which does not seek optimal fusion like MV but suboptimal, robust fusion in tune with the heavy-tailed estimator. 
To this purpose, we employ the arithmetic average (AA) density fusion that basically merges the involved posteriors in an appropriate way. The AA fusion, often applied jointly with the consensus/diffusion approach 
and finite mixture optimization, was first proposed for multi-target density fusion in the presence of missing and false data \cite{Li17PC,Li17PCsmc,Kim20_5G,Ramachandran21,Li20AAmb,Li21AApmbm} 
and has recently been extended to a single target \cite{Li19Bernoulli,Zheng22} and to the classic KF \cite{Li21some}; see also cutting-edge reviews \cite{Da21Recent,Koliander22}. It has demonstrated high efficiency in computation and tolerance to sensor fault (such as misdetection and false alarm), and accommodates any degree of inter-sensor correlation. More relevantly, the AA density fusion is rooted in a dispersive finite mixture expression of the target distribution and so is naturally robust to outlier and complies with the principle for heavy-tailed estimator design. 

However, the straightforward application of the AA fusion to the Student's $t$ distribution leads to a Student's $t$ mixture which 
has to be approximated by a single $t$ distribution for the purpose of closed-formed recursive filtering. Furthermore, the optimization of the fusion weights drives a need of Kullback-Leibler (KL) divergence regarding $t$ distributions which does not admit analytical solution. To solves these challenges, we resort to ideas in dealing with Gaussian-AA fusion \cite{Li21some} through approximating calculation regarding $t$ distributions by that of Gaussian distributions with matched mean and covariance. To concentrate our key contribution on Student's $t$ AA fusion, the single sensor filter we employ is the Student's $t$ filter based on simplified dof choice \cite{Roth13,Roth17}. All derivations are made from first principles. As an extension of the proposed multi-sensor Student's $t$ filter, the covariance intersection (CI) \cite{Julier01} 
approach advocated originally for Gaussian fusion can also be employed in place of the AA fusion. 

%


The remainder of this paper is organized as follows. Background 
is briefly introduced in section \ref{sec:background}. The proposed Student's $t$ filter based on the AA fusion approach is driven in section \ref{sec:t-averaging} and is extended to the CI fusion in section \ref{sec:extention}. Simulation and comparison study are given in section \ref{sec:simulation} before we conclude in section \ref{sec:conclusion}.

\section{Background} \label{sec:background}
\subsection{State Space Model with Student's $t$ Noises}
We consider the following discrete-time state space model with additive noises
\begin{align}
\mathbf{x}_{k} & = {f}_{k-1}(\mathbf{x}_{k-1}) + \mathbf{w}_{k-1} \\
\mathbf{z}_{k} & = {h}_{k}(\mathbf{x}_{k}) + \mathbf{v}_{k}
\end{align}
where ${k}$ is the discrete time index, $\mathbf{x}_k\in\mathbb{R}^{n}$ is the state vector, $\mathbf{z}_k\in\mathbb{R}^{m}$ is the measurement vector, and ${f}_{k-1}(\cdot)$ and ${h}_{k}(\cdot)$ are the process and measurement functions, respectively, and $\mathbf{w}_{k}$ and $\mathbf{v}_{k}$ are the process and measurement noises, respectively. We further denote the  Jacobian matrixes of ${f}_{k-1}(\cdot)$ and ${h}_{k}(\cdot)$ by $\mathbf{F}_{k-1}$ and $\mathbf{H}_{k}$, respectively, which are useful for linearizing the nonlinear models. 

In this paper, we particularly consider the heavy-tailed Student's $t$ noises, namely, 
\begin{align}
p(\mathbf{w}_{k})  &= \mathcal{S}(\mathbf{w}_{k};\mathbf{0},\mathbf{Q}_{k},\nu_{Q}) \\
p(\mathbf{v}_{k})  &= \mathcal{S}(\mathbf{v}_{k};\mathbf{0},\mathbf{R}_{k},\nu_{R})
\end{align}
where $\mathcal{S}(\cdot;\mathbf{\mu},\mathbf{\Sigma},\nu)$ denotes the Student's $t$ PDF with mean vector $\mathbf{\mu}$, scale matrix $\mathbf{\Sigma}$, and degree of freedom (dof) parameter $\nu$, $\mathbf{Q}_{k}$ and $\nu_{Q}$ are the scale matrix and dof parameter of the process noise, respectively. 

The Student's $t$ distribution $\mathcal{S}(\mathbf{x} ; \mathbf{\mu}, \mathbf{\Sigma}, \nu)$ is a bell-shaped distribution with PDF \cite{Nadarajah05}
\begin{align}
& \mathcal{S}(\mathbf{x}; \mathbf{\mu}, \mathbf{\Sigma}, \nu) \nonumber \\
& = \frac{\Gamma(\frac{\nu+n}{2})}{\Gamma(\frac{\nu}{2})|\Sigma|^{\frac{1}{2}}(\pi\nu)^{\frac{d}{2}}}\left( 1+\frac{(\mathbf{x}-\mathbf{\mu})^T\mathbf{\Sigma}^{-1}(\mathbf{x}-\mathbf{\mu})}{\nu}\right)^{-\frac{\nu+n}{2}}  \label{eq:Stu-t-pdf}
\end{align}
where $\Gamma(\cdot)$ represents the Gamma function.
To note, the mean and covariance of the $\mathcal{S}(\mathbf{x} ; \mathbf{\mu}, \mathbf{\Sigma}, \nu)$ are $\mathbf{\mu}$ and $\frac{\nu}{\nu-2} \mathbf{\Sigma}$, respectively. Hereafter, $\nu>2$.

\subsection{Student's $t$ Recursion}
In what follows, the initial state vector $\mathbf{x}_{0}$ is assumed to have a Student's $t$ distribution, namely $p(\mathbf{x}_{0}) = \mathcal{S}(\mathbf{x};\mathbf{\hat{x}}_{0},\mathbf{P}_{0},\nu_{0})$, and $\mathbf{x}_{0}$, $\mathbf{w}_{k}$ and $\mathbf{v}_{k}$ are assumed to be mutually uncorrelated.
Assume the joint distribution of the state and process noise as Student's $t$ with joint dof $\nu'_k$ and parameters $\mathbf{P}'_{k}$ and $\mathbf{Q}'_{k}$, namely,
\begin{equation}\label{eq:joint-xw}
p(\mathbf{x}_{k}, \mathbf{w}_{k}\vert \mathbf{Z}_{1:k}) \nonumber \\
= \mathcal{S} \bigg(\begin{bmatrix} \mathbf{x}_{k} \\ \mathbf{v}_{k-1} \end{bmatrix}; \begin{bmatrix} \hat{\mathbf{x}}_{ k} \\
   0 \end{bmatrix}, \begin{bmatrix} \mathbf{P}'_{k}  \quad 0\\ 0 \quad \mathbf{Q}'_{k} \end{bmatrix}, \nu'_k\bigg)
\end{equation}
then from the rules for linear transformation of $t$ vectors \cite{Roth17}, one gets
\begin{align}
p(\mathbf{x}_{k},  & \mathbf{x}_{k+1}\vert \mathbf{Z}_{1:k+1}) \nonumber \\
& = \mathcal{S} \bigg(\begin{bmatrix} \mathbf{x}_{k} \\ \mathbf{x}_{k+1} \end{bmatrix}; \begin{bmatrix} \hat{\mathbf{x}}_{k} \\
   \hat{\mathbf{x}}_{k+1}  \end{bmatrix}, \begin{bmatrix} \mathbf{P}'_{k}  \quad \quad \mathbf{F}_{k}\mathbf{P}'_{k}\\\mathbf{P}'_{k}\mathbf{F}_{k}^\text{T} \quad \mathbf{P}_{k+1|k} \end{bmatrix}, \nu'_k\bigg)  \label{eq:joint-xx}
\end{align}
The choices of parameters $\nu'$, $\mathbf{P}'_{k}$ and $\mathbf{Q}'_{k}$ are discussed in \cite{Roth17}. In this work, we use the simple choice that $\nu'_k=\min(\nu_k,\nu_{Q})$, $\mathbf{P}'_{k}=\mathbf{P}_{k}$ and $\mathbf{Q}'_{k} = \mathbf{Q}_{k}$. The one-step state prediction results in a $t$ density
\begin{equation}\label{eq:Stu-t-predictoon}
  p(\mathbf{x}_{k+1} \vert \mathbf{Z}_{1:k}) = \mathcal{S}(\mathbf{x}_{k+1};\mathbf{\hat{x}}_{{k+1|k}},\mathbf{P}_{{k+1|k}},\nu'_k)
\end{equation}
where $\mathbf{\hat{x}}_{{k+1|k}} = {f}_{k}(\mathbf{\hat{x}}_{k})$ and $\mathbf{P}_{{k+1|k}} = \mathbf{F}_{k}\mathbf{P}'_{k} \mathbf{F}_{k}^\text{T} + \mathbf{Q}'_{k}$.

Similarly, assume the joint distribution of the predicted state and measurement noise as Student's $t$ with joint dof $\nu'_{k+1}$ and parameters $\mathbf{P}'_{k+1|k}$ and $\mathbf{R}'_{k+1|k}$, namely,
\begin{align} 
p(& \mathbf{x}_{k+1}, \mathbf{v}_{k+1}\vert \mathbf{Z}_{1:k}) \nonumber \\
 &= \mathcal{S} \bigg(\begin{bmatrix} \mathbf{x}_{k+1} \\ \mathbf{v}_{k+1} \end{bmatrix}; \begin{bmatrix} \hat{\mathbf{x}}_{k+1|k} \\
   0 \end{bmatrix}, \begin{bmatrix} \mathbf{P}'_{k+1|k}  \quad \quad 0\\ 0 \quad\quad \mathbf{R}'_{k+1} \end{bmatrix}, \nu'_{k+1}\bigg) \label{eq:joint-x-v}
\end{align}
Again, a simple choice is $\nu'_{k+1}=\min(\nu'_{k},\nu_{R})$, $\mathbf{P}'_{k+1|k}=\mathbf{P}_{k+1|k}$ and $\mathbf{R}'_{k+1} = \mathbf{R}_{k+1}$. Consequently, the prediction density of the state and output can be written as
\begin{align}
& p(\mathbf{x}_{k+1}, \mathbf{z}_{k+1}\vert \mathbf{Z}_{1:k}) \nonumber \\
& = \mathcal{S} \bigg(\begin{bmatrix} \mathbf{x}_{k+1} \\ \mathbf{z}_{k+1} \end{bmatrix}; \begin{bmatrix} \hat{\mathbf{x}}_{k+1|k} \\
   \hat{\mathbf{z}}_{k+1}  \end{bmatrix}, \begin{bmatrix} \mathbf{P}'_{k+1|k}  \quad \mathbf{H}_{k+1}\mathbf{P}'_{k+1|k}\\\mathbf{P}'_{k+1|k}\mathbf{H}_{k+1}^\text{T} \quad \mathbf{S}_{k+1} \end{bmatrix}, \nu'_{k+1}\bigg)  \label{eq:joint-xz}
\end{align}
where $\mathbf{\hat{z}}_{k+1} = {h}_{k+1}(\mathbf{\hat{x}}_{k+1|k})$ and $\mathbf{S}_{{k+1}} = \mathbf{R}'_{k+1} + \mathbf{H}_{k+1}\mathbf{P}'_{k+1|k} \mathbf{H}_{k+1}^\text{T}$,

Then, the final Student's $t$ posterior is given by a $t$ density
\begin{equation}\label{eq:Stu-t-update}
  p(\mathbf{x}_{k+1} \vert \mathbf{Z}_{1:k+1}) = \mathcal{S}(\mathbf{x}_{k+1};\mathbf{\hat{x}}_{{k+1}},\mathbf{P}_{{k+1}},\nu_{k+1})
\end{equation}
where
\begin{align}
\mathbf{\hat{x}}_{{k+1}} & = \mathbf{\hat{x}}_{k+1|k}+\mathbf{K}_{k+1}(\mathbf{z}_{k+1}-\mathbf{\hat{z}}_{k+1}) \\
\mathbf{P}_{{k+1}}  & = \alpha_{k+1} (\mathbf{P}_{k+1|k}-\mathbf{K}_{k+1}\mathbf{S}_{k+1}\mathbf{K}_{k+1}^\text{T}) \\  
\nu_{k+1} & = \nu'_{k+1}+m
\end{align}
with
 \begin{align}
 \alpha_{k+1} & = \frac{\nu'_{k+1}+(\mathbf{z}_{k+1}-\mathbf{\hat{z}}_{k+1})^\text{T}\mathbf{S}_{k+1}^{-1}(\mathbf{z}_{k+1}-\mathbf{\hat{z}}_{k+1})}{(\nu'_{k+1}+m)} \\
\mathbf{K}_{k+1} & = \mathbf{P}'_{k+1|k}\mathbf{H}_{k+1}^\text{T}\mathbf{S}_{k+1}^{-1}
\end{align}

Simply, when $\nu_0=\nu_{Q}=\nu_{R}$, the above Student's $t$ recursion differs from the classical KF merely in the calculation of the scale matrix. To highlight this similarity, we refer to this basic procedure as Student's $t$ Kalman filter (StKF), which is the backbone of our proposed multi-sensor AA fusion StKF. The Monte Carlo method \cite{Genz98}, linearization \cite{Roth13}, and unscented transform \cite{Huang16,Tronarp16} 
can be the same applied for nonlinear StKF as in the nonlinear KFs.

\subsection{Suboptimal AA Density Fusion}
Whenever optimal multi-sensor fusion is sought, the exact correlation between these sensors is needed. 
Unfortunately, despite ideal cases with a-priori information \cite{Bar-Shalom86,Chong04}, 
it is practically intractable to do so. When a sensor network is involved, it becomes even more challenging as the correlation between more significant and complicated.
Then, one may resort to suboptimal, {robust fusion} which eschews underestimating the actual squared estimate errors \cite{Uhlmann96,Julier01,Bailey12} and gains robustness. The AA fusion method is one of the fusion methods such motivated at first \cite{Bailey12}. In fact, the AA fusion has a variety of important statistic and information-theoretic properties such as the capacity to preserve modes in the fusing sources and to combat false alarm in addition to its high computation efficiency; the reader is kindly referred to \cite{Li19Second,Da21Recent,Li21some,Koliander22} for overviews of this method.

Straightforwardly, given a number of probability distributions ${f_i}(\mathbf{x})$ (of the same family or not)
yielded by different estimators $i\in \mathcal{I} :=\{1,2,\cdots,I \in \mathbb{N}^+\}$, the AA fusion approach approximates the target distribution $g(\mathbf{x})$ by their weighted AA
\begin{equation}\label{eq:defAA}
  f_\text{AA}(\mathbf{x}) = \sum_{i \in \mathcal{I}} {w_i}f_i(\mathbf{x})
\end{equation}
where $\mathbf{w} \in \mathbb{W}:=\{\mathbf{w} \in \mathbb{R}^{I}|\mathbf{w}^\mathrm{T}\mathbf{1}_I = 1, w_i > 0, \forall i \in \mathcal{I}\} \subset \mathbb{R}^{I}$ are positive, normalized fusing weights. 

The mixture distribution produced by the AA density fusion is conducive to recursive filtering calculation in two aspects: First, a mixture of conjugate priors is also conjugate and can approximate any kind of prior \cite{Dalal83mixturePrior,Diaconis83mixturePrior}. Second, the linear fusion of a finite number of mixtures of the same parametric family remains a mixture of the same family \cite{Li21AApmbm,Da21Recent}. Third, for any target density $g(\mathbf{x})$, the AA ensures a better fit on average than those fusing densities $D_\text{KL}\left({f_\text{AA}}\| g\right) \leq \sum_{i\in \mathcal{I}} {w_i} D_\text{KL}({f_i}\| g)$, where the equation holds if and only if all densities ${f_i}, i\in \mathcal{I}$ are identical \cite{Blahut87,Li21some}.

Mathematically, the AA fusion minimizes the weighted sum of the KL divergences of the fused result 
with relative to the fusing densities as follows \cite{Abbas09,DaKai_Li_DCAI19} 
\begin{equation}
  f_\text{AA}(\mathbf{x}) = \mathop{\arg\min}\limits_{g \in \mathcal{F}} \sum_{i \in \mathcal{I}}{w_iD_\text{KL}\big(f_i||g\big)} \label{eq:AA_KL divergence} \\
\end{equation}
where $\mathcal{F}$ specifies the goal function space.

The above minimization holds for any fusing weights belong to $\mathbb{W}$. A suboptimal weighting solution, referred to as the \textit{diversity preference} solution \cite{Li21some}, is given as
\begin{equation}
  \mathbf{w}_\text{subopt} = \mathop{\arg\max}\limits_{\mathbf{w} \in \mathbb{W}}  \sum_{i \in \mathcal{I}} w_i D_\text{KL}( {f_i}\| f_\text{AA}) \label{eq:subAAweight}
\end{equation}

Combining \eqref{eq:AA_KL divergence} and \eqref{eq:subAAweight} results in the suboptimal AA fusion as follows \cite{Li21some}
\begin{equation}
(\mathbf{w}_\text{subopt},f_\text{AA}) = 
  \arg\mathop{\max}\limits_{\mathbf{w} \in \mathbb{W}} \mathop{\min}\limits_{g \in \mathcal{F}} \sum_{i \in \mathcal{I}} w_i D_\text{KL}( {f_i}\| g)  \label{eq:JiontOpt-AA}
\end{equation}

\section{Proposed Multi-sensor StKF based on AA Density Fusion} \label{sec:t-averaging}
In this section, we first highlight an important statistical property of the AA density fusion. This property facilitates approximating the AA of Student's $t$ densities by a single $t$ distribution. The suboptimal fusing weights are driven from the information-theoretic optimization \eqref{eq:JiontOpt-AA} via moment-matched Gaussian
approximation to $t$ distribution, leading to the suboptimal AA fusion based multi-sensor StKF.

\subsection{Statistics of AA Density Fusion}
With regard to a finite mixture, there are two important properties:
First, the moments of the mixture are the convex combination of those of the sub-posteriors, i.e.,
\begin{equation}\label{eq:AA_mean}
  \mathbb{E}_{f_\text{AA}}\left[\mathbf{x}_k\right] = \sum_{i \in \mathcal{I}} {{w_i}\mathbb{E}_{f_i}\left[\mathbf{x}_k\right]},
\end{equation}
where $\mathbb{E}_{f}[\mathbf{x}]$ denotes any moment of the variable $\mathbf{x}$ with distribution ${f}(\mathbf{x})$ and \eqref{eq:AA_mean} can be easily seen from the definition of the $n$-th order moment $\mathbb{E}^n_{f}[\mathbf{x}] \triangleq \int \mathbf{x}^n {f}(\mathbf{x}) d\mathbf{x}$. In particular, on the first order moment $n=1$ which is the mean of the density, i.e., $\mathbf{\hat{x}}_i \triangleq \int \mathbf{x} {f}_i(\mathbf{x}) d\mathbf{x}$, we have $\mathbf{\hat{x}}_\text{AA} = \sum_{i \in \mathcal{I}} {w_i} {\mathbf{\hat{x}}_i}$.

Second, the variance of the mixture is driven as
\begin{align}
  \mathbf{P}_{f_\text{AA}}  & \triangleq  \int \left(\mathbf{x}-\mathbb{E}_{f_\text{AA}}\left[\mathbf{x}_k\right]\right)\left( \right)^\text{T} {f}_\text{AA}(\mathbf{x}) d\mathbf{x} \nonumber \\
  & =  \sum_{i \in \mathcal{I}} {w_i} \int \left(\mathbf{x}-\mathbb{E}_{f_\text{AA}}\left[\mathbf{x}_k\right]\right)\left( \right)^\text{T} f_i(\mathbf{x}) d\mathbf{x}  \nonumber \\
  & =  \sum_{i \in \mathcal{I}} {w_i} \int \left(\mathbf{x}-\mathbb{E}_{f_i}\left[\mathbf{x}_k\right]+ \mathbf{\tilde{x}}_i \right)\left( \right)^\text{T} f_i(\mathbf{x}) d\mathbf{x}  \nonumber \\
  & =  \sum_{i \in \mathcal{I}} {w_i} \left(\int \left(\mathbf{x}-\mathbb{E}_{f_i}\left[\mathbf{x}_k\right] \right)\left( \right)^\text{T} f_i(\mathbf{x}) d\mathbf{x} + \mathbf{\tilde{x}}_i \mathbf{\tilde{x}}^\mathrm{T}_i \right) \nonumber \\
  & =  \sum_{i \in \mathcal{I}} {{w_i} \Big(\mathbf{P}_{f_i} + \mathbf{\tilde{x}}_i \mathbf{\tilde{x}}^\mathrm{T}_i\Big)} \label{eq:AA_Variance} 
\end{align}
where $ \mathbf{\tilde{x}}_i := \mathbb{E}_{f_i}\left[\mathbf{x}_k\right] - \mathbb{E}_{f_\text{AA}}\left[\mathbf{x}_k\right] $ and $(\mathbf{a})()^\text{T} := (\mathbf{a})(\mathbf{a})^\text{T}$. 

%


\subsection{Averaging Student's $t$ Densities}
Consider the AA of $t$ densities $\mathcal{S}_i(\mathbf{x}_{k};\mathbf{\hat{x}}_{i,k},\mathbf{P}_{i,k},\nu_{i,k}) , i \in \mathcal{I}$ 
\begin{equation}\label{eq:AA-t-mixture}
  f^{\mathcal{S}}_\text{AA}(\mathbf{x}_k) = \sum_{i \in \mathcal{I}} {w_i}\mathcal{S}_i(\mathbf{x}_{k};\mathbf{\hat{x}}_{i,k},\mathbf{P}_{i,k},\nu_{i,k})
\end{equation}

According to \eqref{eq:AA_mean} and \eqref{eq:AA_Variance}, we have
\begin{align}\label{AA-t-moments}
  \mathbb{E}_{f^{\mathcal{S}}_\text{AA}}\left[\mathbf{x}_k\right]  & = \sum_{i \in \mathcal{I}} {{w_i}\mathbb{E}_{{\mathcal{S}}_i}\left[\mathbf{x}_k\right]} \nonumber \\
  & = \sum_{i \in \mathcal{I}} {w_i} \mathbf{\hat{x}}_{i,k} \\
  \mathbf{P}_{f^{\mathcal{S}}_\text{AA}} & = \sum_{i \in \mathcal{I}} {{w_i} \Big( \mathbf{P}_{{\mathcal{S}}_i} + \mathbf{\tilde{x}}_i \mathbf{\tilde{x}}^\mathrm{T}_i\Big)} \nonumber \\
  & =\sum_{i \in \mathcal{I}} {w_i} \Big( \frac{\nu_{i,k}}{\nu_{i,k}-2}\mathbf{P}_{i,k}  + \big( \mathbf{\hat{x}}_{i,k}-\sum_{i \in \mathcal{I}} {w_i} \mathbf{\hat{x}}_{i,k}\big) \big( \big)^\mathrm{T}\Big) \label{eq:Stu-P-AA}
\end{align}
This leads to our following key result which ensures the closed-from recursion of the StKF, namely the fusion of Student's $t$ posteriors remains $t$ distributed, if the AA fusion is employed for multi-sensor density fusion.
\begin{proposition} The first and second moment matched-based $t$ density approximation to the weighted mixture of Student's $t$ densities as in \eqref{eq:AA-t-mixture} is given by
\begin{align}
  \mathcal{S}_\text{AA}(\mathbf{x}_k) &\approx \mathcal{S}(\mathbf{x}_{k}; \mathbf{\hat{x}}_\text{AA}, \mathbf{P}_\text{AA},\nu_\text{AA}) \label{eq:StuPDF-aa} \\
\mathbf{\hat{x}}_\text{AA} & = \sum_{i \in \mathcal{I}} {w_i} \mathbf{\hat{x}}_{i,k} \label{eq:m-AA} \\
 \mathbf{P}_\text{AA} &= \frac{\nu_\text{AA}-2}{\nu_\text{AA}}\mathbf{P}_{f^{\mathcal{S}}_\text{AA}} \label{eq:P-AA}
\end{align}
where $\mathbf{P}_{f^{\mathcal{S}}_\text{AA}}\left[\mathbf{x}_k\right]$ is given in \eqref{eq:Stu-P-AA} and one may choose $\nu_\text{AA}=\min_{i\in \mathcal{I}}\{\nu_{i,k}\}$ to preserve the heaviest tail of all fusing densities or choose their average $\nu_\text{AA}=\frac{1}{|\mathcal{I}|}\sum_{i\in \mathcal{I}}\nu_{i,k}$. 
\end{proposition}

\subsection{Suboptimal Weight for Averaging Student's $t$ Densities} \label{sec:suboptimalweight}
Substituting Student's $t$ densities and the moment-matched approximation \eqref{eq:StuPDF-aa} into the suboptimal weighting solution \eqref{eq:subAAweight} yields
\begin{align}
  \mathbf{w}_\text{subopt} & = \mathop{\arg\max}\limits_{\mathbf{w} \in \mathbb{W}} \sum_{i \in \mathcal{I}} w_i D_\text{KL}({\mathcal{S}_i}\| f^\mathcal{S}_\text{AA}) \label{eq:subW-t-AA} \\
        & \approx  \mathop{\arg\max}\limits_{\mathbf{w} \in \mathbb{W}}  \sum_{i \in \mathcal{I}} w_i D_\text{KL}( {\mathcal{S}_i}\| \mathcal{S}_\text{AA}) \label{eq:subW-t-t}
\end{align}
Unfortunately, both \eqref{eq:subW-t-AA} and \eqref{eq:subW-t-t} do not admit analytical solution. In spite of numerical approximation methods such as the Monte Carlo method,
we propose a further simplified alternative based on the similarity between the Student's $t$ and Gaussian distributions as follows
\begin{proposition} The KL divergence between Student's $t$ distributions $\mathcal{S}_i(\mathbf{x}) = \mathcal{S}(\mathbf{x};\mathbf{\hat{x}}_i, \frac{\nu_i-2}{\nu_i}\mathbf{P}_i, \nu_i)$ and $\mathcal{S}_j(\mathbf{x}) = \mathcal{S}(\mathbf{x};\mathbf{\hat{x}}_j, \frac{\nu_j-2}{\nu_j}\mathbf{P}_j, \nu_j)$ is approximated as that between their moment-matched Gaussian distributions $\mathcal{N}_i(\mathbf{x}) = \mathcal{N}(\mathbf{x};\mathbf{\hat{x}}_i, \mathbf{P}_i)$ and $\mathcal{N}_j(\mathbf{x}) = \mathcal{N}(\mathbf{x};\mathbf{\hat{x}}_j, \mathbf{P}_j)$.
%
This leads to an analytically approximate weight solution
\begin{align}
  & \mathbf{w}_\text{subopt} \approx \mathop{\arg\max}\limits_{\mathbf{w} \in \mathbb{W}} \sum_{i \in \mathcal{I}} w_i D_\text{KL}( {\mathcal{S}_i}\| \mathcal{S}_\text{AA}) \\
  & \approx \mathop{\arg\max}\limits_{\mathbf{w} \in \mathbb{W}} \sum_{i \in \mathcal{I}} w_i D_\text{KL}\big( {\mathcal{N}_i}({\mathbf{\hat{x}}}_{i},{\mathbf{P}}_{i})\| \mathcal{N}_\text{AA}({\mathbf{\hat{x}}}_\text{AA},{\mathbf{P}}_\text{AA})\big) \\
  & = \mathop{\arg\max}\limits_{\mathbf{w} \in \mathbb{W}} \sum_{i \in \mathcal{I}} w_i\bigg[\frac{\nu_i}{\nu_i-2}{\mathrm {tr}}\big(\mathbf{P}^{-1}_{f^{\mathcal{S}}_\text{AA}}\mathbf{P}_{i}\big) + \log{\det(\mathbf{P}_{f^{\mathcal{S}}_\text{AA}})\over \det(\frac{\nu_i}{\nu_i-2}\mathbf{P}_{i})} \nonumber \\
      & \quad \quad + ({\mathbf{\hat{x}}}_{i}-{\mathbf{\hat{x}}}_\text{AA})^\mathrm{T}\mathbf{P}^{-1}_{f^{\mathcal{S}}_\text{AA}}({\mathbf{\hat{x}}}_{i}-{\mathbf{\hat{x}}}_\text{AA}) \bigg] \label{eq:final-sub-w}
\end{align}
\end{proposition}

There is another alternative approximation rather than the above exact moment matching. That is, ignore the dof of the Student's $t$ distribution and treat its scale matrix as a covariance, i.e., $D_\text{KL}( {\mathcal{S}_i(\mathbf{\hat{x}}_i, \mathbf{P}_i, \nu_i)}\| \mathcal{S}_j(\mathbf{\hat{x}}_j, \mathbf{P}_j, \nu_j)) \approx D_\text{KL}( {\mathcal{N}_i(\mathbf{\hat{x}}_i,  \mathbf{P}_i)}\| \mathcal{N}_j(\mathbf{\hat{x}}_j,\mathbf{P}_j))$. This will lead to a novel analytical weight solution as follows, c.f., \eqref{eq:final-sub-w},
\begin{align}
  & \mathbf{w}_\text{subopt} \nonumber \\
  & \approx  \mathop{\arg\max}\limits_{\mathbf{w} \in \mathbb{W}} \sum_{i \in \mathcal{I}} w_i\bigg[\frac{\nu_i}{\nu_i-2}{\mathrm {tr}}\big(\mathbf{P}^{-1}_{f^{\mathcal{S}}_\text{AA}}\mathbf{P}_{i}\big)  + \log{\det(\frac{\nu_\text{AA}}{\nu_\text{AA}-2}\mathbf{P}_{f^{\mathcal{S}}_\text{AA}})\over \det(\mathbf{P}_{i})}
 \nonumber \\
  & \quad \quad +\frac{\nu_\text{AA}-2}{\nu_\text{AA}}({\mathbf{\hat{x}}}_{i}-{\mathbf{\hat{x}}}_\text{AA})^\mathrm{T}\mathbf{P}^{-1}_{f^{\mathcal{S}}_\text{AA}}({\mathbf{\hat{x}}}_{i}-{\mathbf{\hat{x}}}_\text{AA}) \bigg] \label{eq:final-sub-w-2}
\end{align}

\begin{figure}
\centering
\vspace{-3mm}
\centerline{\includegraphics[width=9.5 cm]{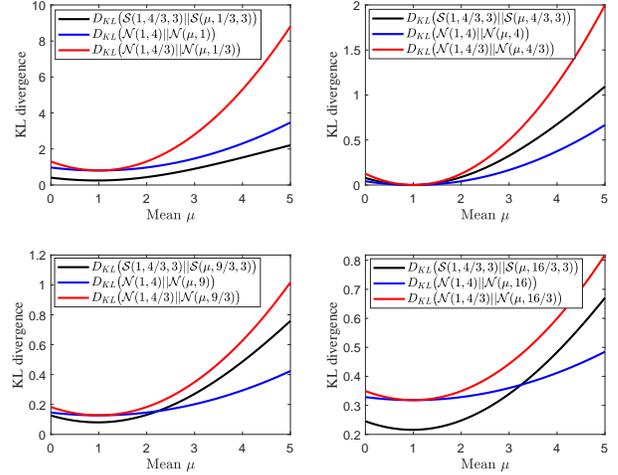}}
\vspace{-1.5mm}
\caption{Gaussian-based approximation to KL divergence between two Student's $t$ densities with dof $\nu =3$.} \label{fig:Stu-Norm-KLD}
\vspace{-1.5mm}
\end{figure}

Obviously, both approximations will converge to the real KL divergence as the dof goes to infinite $\nu\rightarrow +\infty$. To illustrate the approximation accuracy, we consider some examples for $\nu=3$ as given in Fig. \ref{fig:Stu-Norm-KLD}. As shown, both methods achieve the minimum value at the same point when the involved two ($t$ or Gaussian) densities have the same means, from which they both monotonously increase faster or slower, as well as the real KL divergence, with the increase of the distance between the means of two densities. This confirms the consistency of the approximation of both methods that a larger approximate divergence indicates a larger real divergence. To gain more insight, note that the solution to \eqref{eq:subW-t-AA} should satisfy the following `middle distribution' equation \cite{Nielsen13Chernoff,Li21some}
\begin{equation}\label{eq:mid-point}
  D_\text{KL}( {\mathcal{S}_i}\| \mathcal{S}_\text{AA}(\mathbf{w}_\text{subopt})) = D_\text{KL}( {\mathcal{S}_j}\| \mathcal{S}_\text{AA}(\mathbf{w}_\text{subopt}))
\end{equation}

\begin{figure}
\centering
\vspace{-3mm}
\centerline{\includegraphics[width=9.5 cm]{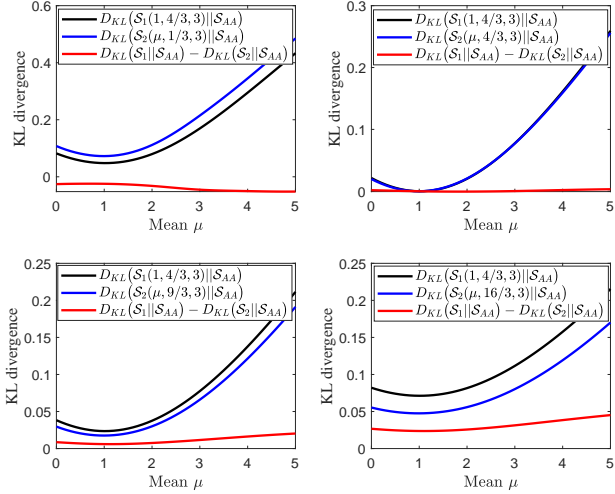}}
\vspace{-1.5mm}
\caption{KL divergences of the fused Student's $t$ with relative to each of the two fusing Student's $t$ densities with dof $\nu =3$ based on the suboptimal fusing weights \eqref{eq:final-sub-w}.} \label{fig:MidKLD}
\vspace{-2mm}
\end{figure}

Therefore, one can evaluate how much the above equation is satisfied to evaluate whether the solutions provided by \eqref{eq:final-sub-w} or \eqref{eq:final-sub-w-2} are accurate. We illustrate here the accuracy of \eqref{eq:final-sub-w} again by examples for $\nu=3$ in Fig. \ref{fig:MidKLD}. It shows that the AA fusion resulted by the proposed suboptimal fusing weights \eqref{eq:final-sub-w} complies well with equation \eqref{eq:mid-point}, especially when the fused distributions have similar covariances. In the case that two Student's $t$ distributions have the same scale matrix (as shown in the right-upper sub-figure of Fig. \ref{fig:MidKLD}), the result meets greatly the `middle distribution' equation \eqref{eq:mid-point} and confirms the effectiveness of \eqref{eq:final-sub-w}. 
This is an important finding: although the Gaussian-$t$ approximation/substitution can be inaccurate, the fusing weights yielded by \eqref{eq:subW-t-t} and \eqref{eq:final-sub-w} can be close with each other and so be gratifying (close to the optimal solution satisfying \eqref{eq:mid-point}).
These being said, more accurate, analytical calculation or approximate method is desirable. 
Approaches such as mode approximation \cite{Loxam08} and variational inference \cite{Zhu13VBStuT,Huang19kld} are notable.

As a final result based on simplified choice of the dof and Gaussian-$t$ approximation, \eqref{eq:final-sub-w} (or \eqref{eq:final-sub-w-2}) and \eqref{eq:StuPDF-aa} constitute the overall information-theoretic optimization-based suboptimal AA fusion of Student's $t$ as follows, c.f. \eqref{eq:JiontOpt-AA}
\begin{equation}
(\mathbf{w}_\text{subopt},\mathcal{S}_\text{AA}) \approx 
  \arg\mathop{\max}\limits_{\mathbf{w} \in \mathbb{W}} \mathop{\min}\limits_{g \in \mathcal{F}} \sum_{i \in \mathcal{I}} w_i D_\text{KL}( {\mathcal{S}_i}\| g)  \label{eq:JiontOpt-Stu-t-AA}
\end{equation}
where $\mathcal{F}(\mathbb{R}_\mathcal{S}^{n_x}) \rightarrow \mathbb{R}$ specifies the Student's $t$ function space.

\subsection{Algorithm Flow of AA-Fusion StKF}
When multiple sensors cooperate with each other, communication will be involved which needs to take into account the inter-sensor connection topology and constraints. There are considerable relevant research, including those for AA fusion \cite{Li17PC,Li17PCsmc,Li19diffusion,Li20AAmb,Zheng22}. However, to avoid distracting the readers' attention, we ignore the communication issue and assume that the sensors have direct access to each other, analogous to the centralized network with feedback. We also assume that the sensors are synchronized and coordinated in the same coordinate system. The overall procedure of the proposed multi-sensor AA-StKF for totally $S$ interconnected sensors is summarized in Algorithm \ref{tab:algorithm}. All parameters of the state space model are assumed known by default.

\begin{algorithm}[h]
	\caption{One Filtering Iteration of Proposed Multi-Sensor AA-Fusion StKF}
	\label{tab:algorithm}
	\begin{algorithmic}[1]
		\REQUIRE $\left \{\mathcal{S}_s(\mathbf{x}_k; \mathbf{\hat{x}}_{s,k}, \mathbf{P}_{s,k}, \nu_{s,k}), \mathbf{z}_{s,k+1} \right \}_{s=1,\cdots,S} $
		\ENSURE  $\left \{ \mathcal{S}_s(\mathbf{x}_{k+1}; \mathbf{\hat{x}}_{s,k+1}, \mathbf{P}_{s,k+1}, \nu_{s,k+1}) \right \}_{s=1,\cdots,S} $
		\FOR {each sensor $s=1,\cdots,S$ in parallel}
        \STATE One-step state prediction as in \eqref{eq:Stu-t-predictoon} in the following steps:
        {\small \begin{enumerate}
          \item $\mathbf{\hat{x}}_{s,{k+1|k}} = {f}_{s,k}(\mathbf{\hat{x}}_{s,k})$,
          \item $\mathbf{P}_{s,{k+1|k}} = \mathbf{F}_{s,k}\mathbf{P}_{s,k} \mathbf{F}_{s,k}^\text{T} + \mathbf{Q}_{s,k}$
          \item $\nu'_{s,k}=\min(\nu_{s,k},\nu_{s,Q})$
        \end{enumerate}
        }
        \STATE Update the prediction as in \eqref{eq:Stu-t-update} using the new measurement $\mathbf{z}_{s,k}$, including the following steps:
        {\small \begin{enumerate}
          \item $\nu'_{s,k+1}=\min(\nu'_{s,k},\nu_{s,R})$
          \item $\mathbf{\tilde{z}}_{s,k+1} = \mathbf{z}_{s,k+1} - {h}_{k+1}(\mathbf{\hat{x}}_{s,k+1|k})$
          \item $\mathbf{S}_{s,{k+1}} = \mathbf{H}_{s,k+1}\mathbf{P}_{s,k+1|k} \mathbf{H}_{s,k+1}^\text{T} + \mathbf{R}_{s,k+1}$
          \item $\alpha_{s,k+1} = \frac{\nu'_{s,k+1}+ \mathbf{\tilde{z}}_{s,k+1}^\text{T}\mathbf{S}_{s,k+1}^{-1}\mathbf{\tilde{z}}_{s,k+1}}{\nu'_{s,k+1}+m_s}$
          \item $\mathbf{K}_{s,k+1} = \mathbf{P}_{s,k+1|k}\mathbf{H}_{s,k+1}^\text{T}\mathbf{S}_{s,k+1}^{-1}$
          \item $\mathbf{\hat{x}}_{s,{k+1}}  = \mathbf{\hat{x}}_{s,k+1|k}+\mathbf{K}_{s,k+1}\mathbf{\tilde{z}}_{s,k+1}$
          \item $\mathbf{P}_{s,{k+1}}  =  \alpha_{s,k+1}(\mathbf{P}_{s,k+1|k}-\mathbf{K}_{s,k+1}\mathbf{S}_{s,k+1}\mathbf{K}_{s,k+1}^\text{T})$
          \item $\nu_{s,k+1} = \nu'_{s,k+1}+m_s$
        \end{enumerate}
        }
		\ENDFOR
		\FOR {each sensor $s=1,\cdots,S$ in parallel}
        \STATE Exchange parameters $\{\mathbf{\hat{x}}_{s,k+1},\mathbf{P}_{s,k+1},\nu_{s,k+1}\}$ among inter-connected sensors.
               Suppose that at the end, sensor $s$ receives $\left \{ \mathcal{S}_i(\mathbf{x}_{k+1}; \mathbf{\hat{x}}_{i,k+1}, \mathbf{P}_{i,k+1}, \nu_{i,k+1}) \right \}_{i \in \mathcal{I}_{s,k}} $. 
		\STATE Calculate the AA fusing weights $\mathbf{w}_\text{subopt}$ via \eqref{eq:final-sub-w} or \eqref{eq:final-sub-w-2}, based on \eqref{eq:Stu-P-AA} and \eqref{eq:m-AA}.
        \STATE Calculate the AA-fused Student's $t$ density as in \eqref{eq:StuPDF-aa} including the following steps:
        {\small \begin{enumerate}
          \item $\nu_{s,k+1} \leftarrow \frac{1}{|\mathcal{I}_{s,k}|}\sum_{i\in \mathcal{I}_{s,k}}\nu_{i,k+1}$ (or $\nu_\text{AA}=\min_{i\in \mathcal{I}_{s,k}}\{\nu_{i,k}\}$)
          \item $\mathbf{\hat{x}}_{s,k+1} \leftarrow \sum_{i \in \mathcal{I}_{s,k}} {w_i} \mathbf{\hat{x}}_{i,k+1}$
          \item $\mathbf{P}_{s,k+1} \leftarrow \frac{\nu_{s,k+1} -2}{\nu_{s,k+1} }\mathbf{P}_{f^{\mathcal{S}}_\text{AA}}\left[\mathbf{x}_{k+1}\right]$ using \eqref{eq:Stu-P-AA}
        \end{enumerate}
        }
		\ENDFOR
 \STATE Steps 5-9 may be performed for multiple iterations in the case of decentralized sensor network, like what is done for average consensus \cite{Li17PC,Li19Bernoulli,Li20AAmb,Zheng22}.
	\end{algorithmic}
\end{algorithm}

\section{Extension of CI Fusion} \label{sec:extention}
As addressed so far, the proposed Student's $t$ AA fusion approach relies on approximating the calculation over $t$ distribution by that over Gaussian. Therefore, many other Gaussian fusion approaches such as the covariance intersection (CI) \cite{Julier01} and inverse CI \cite{Noack17ici} can be applied in place of the AA fusion for multi-sensor Student's $t$ filter design. Here, we briefly address the extension of the CI fusion. The CI fusion of $t$ densities $\mathcal{S}_i(\mathbf{x}_{k};\mathbf{\hat{x}}_{i,k},\mathbf{P}_{i,k},\nu_{i,k}) , i \in \mathcal{I}$ can be written by
\begin{equation}\label{eq:GA-t-fusion}
  f^{\mathcal{S}}_\text{CI}(\mathbf{x}_k) \propto \prod_{i \in \mathcal{I}}\left(\mathcal{S}_i(\mathbf{x}_{k};\mathbf{\hat{x}}_{i,k},\mathbf{P}_{i,k},\nu_{i,k})\right)^{w_i}
\end{equation}

The above fusion is not analytically evolvable due to the non-integrability of the poly-$t$ (product of Student's $t$ distributions) distribution. By using Gaussian distributions $\mathcal{N}_i(\mathbf{x}_{k};\mathbf{\hat{x}}_{i,k}, \frac{\nu_{i,k}}{\nu_{i,k}-2}\mathbf{P}_{i,k})$ to approximate the corresponding Student's $t$ distributions $\mathcal{S}_i(\mathbf{x}_{k};\mathbf{\hat{x}}_{i,k}, \mathbf{P}_{i,k}, \nu_{i,k}), i \in \mathcal{I}$ with the same first and second moments, the CI fusion is given as follows
\begin{align}
\mathbf{\hat{x}}_\mathrm{CI} &=  \mathbf{P}_\mathrm{CI}\sum_{i \in \mathcal{I}} \frac{w_i (\nu_{i,k}-2)}{\nu_{i,k}}\mathbf{P}_{i,k}^{-1} \mathbf{\hat{x}}_i \label{eq:CI-x}\\
\mathbf{P}_\mathrm{CI} &= \bigg(\sum_{i \in \mathcal{I}} \frac{w_i (\nu_{i,k}-2)}{\nu_{i,k}} \mathbf{P}_{i,k}^{-1}\bigg)^{-1} \label{eq:CI-P}
\end{align}
where
\begin{equation}
  \mathbf{w}_\mathrm{CI} =  \mathop{\arg\min}\limits_{\mathbf{w} \in \mathbb{W}} \mathrm{Tr}(\mathbf{P}_\mathrm{CI}) \label{eq:CI-w}
\end{equation}

With regard to Algorithm \ref{tab:algorithm}, the above formulation is amount to using \eqref{eq:CI-w}, \eqref{eq:CI-x} and \eqref{eq:CI-P} to replace \eqref{eq:final-sub-w}/\eqref{eq:final-sub-w-2} in Step 7, Step 8.2) and Step 8.3), respectively. This leads to a multi-sensor CI fusion-based StKF. 

\section{Simulations} \label{sec:simulation}
\label{sec:simulation}

We consider a single target tracking problem. 
Following the literature \cite{Roth13,Huang16}, we simulate abnormal noises of a significant magnitude (much higher than the normal case) which randomly occur with probabilities to assume outliers affecting the state process and the measurement, independently. 
The state of the target $\mathbf{x}_k=[p_{x,k},\dot{p}_{x,k},p_{y,k},\dot{p}_{y,k}]^\mathrm{T}$ consists of planar position $[p_{x,k},p_{y,k}]^\mathrm{T}$ and velocity $[\dot{p}_{x,k},\dot{p}_{y,k}]^\mathrm{T}$.
At time $k=0$, it is randomly initialized as $\mathbf{x}_0 \sim \mathcal{N}(\mathbf{x};\mathbf{\mu}_0,\mathbf{P}_0)$,
where $\mathbf{\mu}_0 = [1000\textrm{m},20\textrm{m}/\textrm{s},1000\textrm{m},0\textrm{m}/\textrm{s}]^\mathrm{T}$ with $\mathbf{P}_0 =$ diag$\{[500\textrm{m}^2,50\textrm{m}^2/\textrm{s}^2,500\textrm{m}^2,50\textrm{m}^2/\textrm{s}^2]\}$, where diag$\{\mathbf{a}\}$ represents a diagonal matrix with diagonal $\mathbf{a}$.
The target moves following a nearly constant velocity motion given as (with the sampling interval $\Delta =1$s)
\begin{equation}\label{eq:Simu_TargetDynamic}
\mathbf{x}_k= \left[ \begin{array}{cccc}
1 & \Delta & 0 & 0 \\
0 & 1 & 0 & 0 \\
0 & 0 & 1 & \Delta \\
0 & 0 & 0 & 1 \\
\end{array} \right] \mathbf{x}_{k-1}+ \left[ \begin{array}{cc}
\frac{\Delta^2}{2} & 0 \\
\Delta & 0 \\
0 & \frac{\Delta^2}{2} \\
0 & \Delta \\
\end{array} \right] \mathbf{u}_{k-1} \ist
\end{equation}
where the process noise $\mathbf{u}_k \sim \mathcal{N}(\mathbf{u};\mathbf{0}_2\textrm{m}/\textrm{s}^2,r^2\mathbf{I}_2\textrm{m}^2/\textrm{s}^4)$. Here, the process noise standard deviation $r$ is defined with the outlier probability $p_o$ as follows
\begin{align}\label{eq:procee-noise}
\bigg\{ \begin{array}{l}
r = 5, \text{with probability $1 - p_o$}  \\
\vspace{0.5mm}
r = 50, \text{with probability $p_o$}
\end{array}
\end{align}

We consider only two sensors. Both sensors $s=1,2$ have the linear measurement model $\mathbf{z}_{s,k}= \mathbf{H}_{s,k} \mathbf{x}_k + \mathbf{v}_{s,k}$ as follows
\begin{equation}\label{eq:linear-measurement-model}
\mathbf{H}_{s,k} =\left[ \begin{array}{cccc}
1  & 0  & 0  & 0 \\
0  & 0  & 1  & 0 \\
\end{array} \right], \mathbf{v}_{s,k} = \left[ \begin{array}{c}
v_{k,1} \\
v_{k,2} \\
\end{array} \right] \ist
\end{equation}
with $v_{k,1}$ and $v_{k,2}$ as mutually independent zero-mean Gaussian noise with the same standard deviation $R_{s}$.

The measurement noise standard deviations of two sensors are defined with outlier probability too as follows (independent with that of the process noise and with each other sensor):
  \begin{align}\label{eq:procee-noise}
  &\bigg\{ \begin{array}{l}
R_{1} \!=\! 20  \mathrm{m}, \text{with probability $1-p_o$}  \\
\vspace{0.5mm}
R_{1}  \!=\! 200  \mathrm{m}, \text{with probability $p_o$}
\end{array} \\
     &\bigg\{ \begin{array}{l}
R_{2} \!=\! 10  \mathrm{m},  \text{with probability $1-p_o$}  \\
\vspace{0.5mm}
R_{2} \!=\! 100  \mathrm{m},   \text{with probability $p_o$}
\end{array}
  \end{align}

We test different outlier probabilities $p_o$ from $0$ (when there is no outlier) to $0.2$. 
In each case of $p_o$, the simulation is performed for 1000 Monte Carlo runs, each having 100 filtering steps. The target trajectory is randomly generated according to the process model with an random initial state in each run. The process and measurement noises are approximated by Student's $t$ noise with the normal mean and covariance/scale parameters as specified, and with dof $\nu_{Q}=\nu_{R}=3$. The root mean square error (RMSE) of the position or velocity estimates is used for filter evaluation
\begin{equation}\label{eq:rmse}
  \mathrm{RMSE}_k= \sqrt{\frac{1}{M}\sum_{i=1}^{M}(\mathbf{\hat{x}}^i_k-\mathbf{x}^i_k)^\text{T}(\mathbf{\hat{x}}^i_k-\mathbf{x}^i_k)}
\end{equation}
where $\mathbf{\hat{x}}^i_k$ is the position or velocity estimate of the real state $\mathbf{{x}}^i_k$ at time $k$ in run $i$ and $M$ is the number of runs.

Both StKF and KF are simulated. The KFs are initialized by $\mathcal{N}(\mathbf{x};\mathbf{\mu}_0,\mathbf{P}_0)$ and the StKFs are initialized by $\mathcal{S}(\mathbf{x};\mathbf{\mu}_0, \frac{\nu_0-2}{\nu_0}\mathbf{P}_0, \nu_0=3)$. Both StKFs and KFs are implemented in both noncooperative manner (using only sensor 1's measurement) and two-sensor cooperative manner. In the latter, we compare the proposed AA, CI fusion with the augmented measurement (AM) approach \cite{Zhu13VBStuT} (a.k.s. centralized batch fusion \cite{Yan21chapter}). In the AM approach, two sensors' measurements are cascaded/augmented as a joint $\mathbf{z}_{k}= \mathbf{H}_{k} \mathbf{x}_k + \mathbf{v}_{k}$ as follows:
\begin{equation}\label{eq:augment-measurement-model}
\mathbf{z}_{k} =\left[ \begin{array}{c}
\mathbf{z}_{1,k} \\
\mathbf{z}_{2,k} \\
\end{array} \right],
\mathbf{H}_{k} =\left[ \begin{array}{c}
\mathbf{H}_{1,k} \\
\mathbf{H}_{2,k} \\
\end{array} \right],
\mathbf{v}_{k} = \left[ \begin{array}{c}
\mathbf{v}_{1,k}  \\
\mathbf{v}_{2,k}  \\
\end{array} \right] \ist
\end{equation}

In addition, the AA fusion has been implemented in two means, one using uniform weights ($w_1=w_2=0.5$) referred to as `unAA fusion' and the other using the suboptimal weight (referred to as AA fusion by default) as given in \eqref{eq:final-sub-w} in section \ref{sec:suboptimalweight}. In both cases, the AA fusion of two Gaussian PDFs results in a Gaussian mixture of two components for which a merging scheme based on moment matching is needed 
to maintain closed-form KF recursion.

\begin{figure}
\centering
\centerline{\includegraphics[width=8cm]{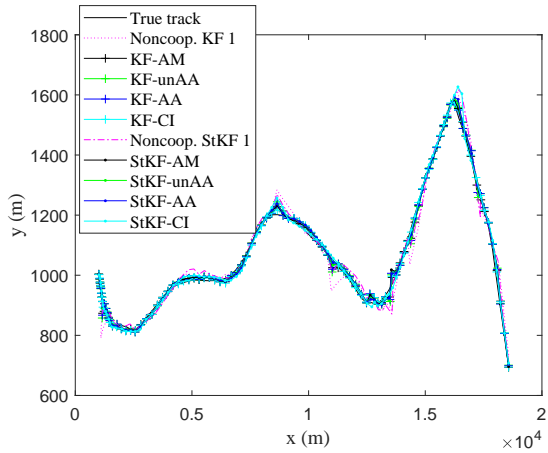}}
\caption{The real target trajectory and estimates of noncooperative and two-sensor-fusion KFs/StKFs in one trial using $p_o=0.05$} \label{fig:Trajectory_Linear}
\end{figure}

\begin{figure}
\centering
\centerline{\includegraphics[width=8cm]{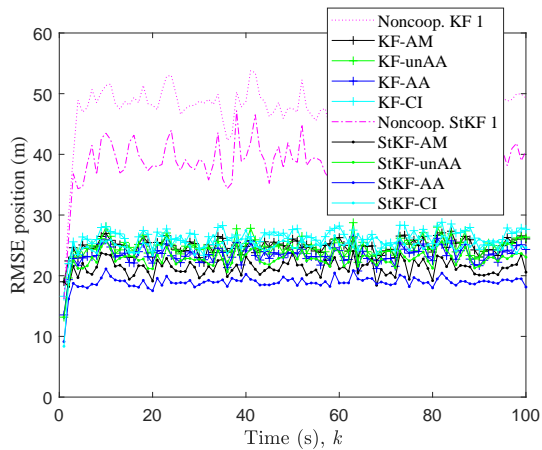}}
\caption{The position RMSEs of noncooperative and two-sensor-fusion KFs/StKFs when $p_o=0.05$} \label{fig:posRMSE_Linear}
\end{figure}

\begin{figure}
\centering
\centerline{\includegraphics[width=8 cm]{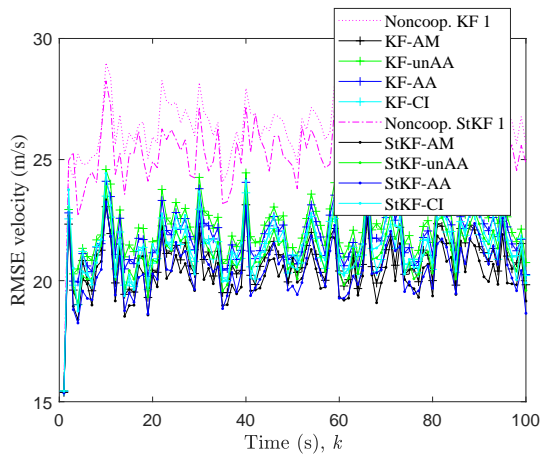}}
\caption{The velocity RMSEs of noncooperative and two-sensor-fusion KFs/StKFs when $p_o=0.05$} \label{fig:velRMSE_Linear}
\end{figure}

\begin{figure}
\centering
\centerline{\includegraphics[width=8 cm]{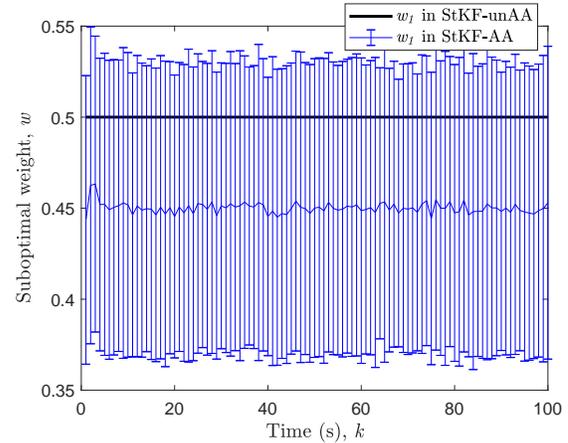}}
\caption{Variance of the fusion weight assigned to sensor 1 in StKF-AA when $p_o=0.05$} \label{fig:fw}
\end{figure}

\begin{figure}
\centering
\centerline{\includegraphics[width=8cm]{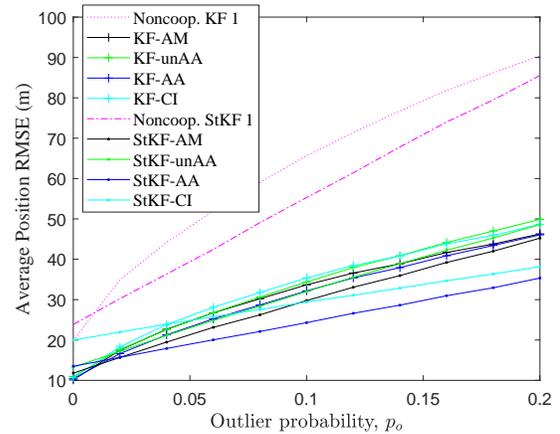}}
\caption{Average position RMSEs against outlier probability} \label{fig:AverposRMSE}
\end{figure}

\begin{figure}
\centering
\centerline{\includegraphics[width=8 cm]{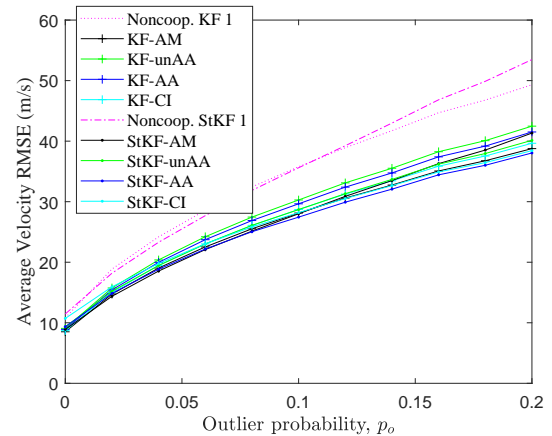}}
\caption{Average velocity RMSEs against outlier probability} \label{fig:AvervelRMSE}
\end{figure}

The real target trajectory and the estimates of the KFs/StKFs in one trial are given in Fig. \ref{fig:Trajectory_Linear}. The position and velocity RMSEs of these filters are given in Fig. \ref{fig:posRMSE_Linear} and Fig. \ref{fig:velRMSE_Linear}, respectively. The average position and velocity RMSEs over all filtering times against outlier probability are given in Fig. \ref{fig:AverposRMSE} and \ref{fig:AvervelRMSE}, respectively. First of all, all fusion methods gain increased estimation accuracy as compared with the noncooperative filters, indicating that these fusion approaches are effective. More specifically, there are several main findings:
\begin{enumerate}
  \item The StKF-AA using the suboptimal fusing weights \eqref{eq:final-sub-w} performs the best in position estimation accuracy as long as $p_o > 0.03$ and in velocity estimation accuracy as long as $p_o \geq 0.08$. Further on, the variance of the fusion weight assigned to sensor 1 in StKF-AA is given in Fig. \ref{fig:fw}, with an average level about $0.45$. This is reasonable since sensor 1 has a larger measurement noise covariance as compared with sensor 2. Indeed, the default AA fusion using optimized weights performs better than that using uniform weights and also the CI fusion, based on whether KF or StKF.
  \item Most StKFs outperform their corresponding KFs using the same fusion approach in position accuracy when $p_o \geq 0.02$ except for the StKF-CI that performs better than KF-CI only when $p_o > 0.04$. This demonstrates the advantage of the Student's $t$ filter in dealing with outlier/heavy-tailed noises as compared with Gaussian filters. In contrast, when there is no outlier ($p_o =0$), the KFs outperform the StKFs, regardless of the fusion approach. This is simply because there is no need to maintain a heavy-tailed posterior for a purely linear Gaussian system.
  \item When there is no outlier, the KF-AM is optimal for the linear Gaussian system. When there exist outliers, the AM approach performs worse than the AA fusion in position accuracy, using whether KF or StKF. The reasons are twofold: First, the AM approach that seeks MV becomes disadvantageous in dealing with outliers. Second, when the outlier is independent across sensors, the AM approach suffers from a higher actual measurement outlier probability which is calculated by
      \begin{equation}\label{eq:AMoutliterProb}
        1-\prod_{i \in \mathcal{I}}(1-p^i_{o}) > \max_{i \in \mathcal{I}} p^i_{o}
      \end{equation}
      where $p^i_{o}$ is the outlier probability of sensor $i$.
\end{enumerate}

 %

Two final remarks are to be highlighted. First, the performance of the basic StKFs using a simplified choice of the dof as addressed is expected to be improved if the dof can (online) adapt with the probability of outlier. Notably, a recent work proposes to combine Gaussian and $t$ on the base of a probabilistic framework so that to adapt the tail \cite{Huang19G-S}. This remains a valuable, open problem. Second, 
the 
CI fusion seeks MV yet covariance consistent fusion which does not have as good fault-tolerant capacity as the AA fusion does \cite{Li21some} and so the forced Gaussian-$t$ approximation may not suit it. As shown, when the outlier probability is small ($p_o < 0.04$), the CI-StKF is disappointing although it is good when $p_o > 0.1$. 
It remains open how to optimally apply Gaussian fusion approaches to the Student's $t$ distributions. 

\section{Conclusion} \label{sec:conclusion}

This paper proposes a multi-sensor Student's $t$ filter based on the AA density fusion approach. An information-theoretic optimization formulation based on moment-matched Gaussian-$t$ approximation is used to drive the suboptimal fusion weights and the $t$ distribution merging procedure to ensure closed-form Student's $t$ recursion. The fusion framework accommodates any Gaussian fusion approaches such as the CI. Simulation based on target tracking using various probabilities of outlier has demonstrated the promising performance of the proposed multi-sensor AA fusion-based $t$ filter in dealing with outliers as compared with the multi-sensor KFs/StKF based on either augmented measurement or covariance intersection.

This work, however, is limited to simplified choice of the dof of the Student's $t$ PDF and inaccurate yet reasonable Gaussian-$t$ approximation for closed-form $t$-AA fusion and for fast and analytical fusing weight design. Improvement can be expected if these limitations are removed or reduced. In particular, a valuable future research is to online adapt the dof of the $t$ density according to the probability (and even the expected magnitude) of outliers.

\bibliographystyle{IEEEtran}
\bibliography{Student}

\end{document}